\documentclass{mn2e}

\usepackage{amssymb}
\usepackage{amsmath}
\usepackage{epsfig}
\title{The Quasar Mass-Luminosity Plane II: High Mass Turnoff Evolution and a Synchronization Puzzle}
\author[Charles L. Steinhardt and Martin Elvis]
       {Charles L. Steinhardt and Martin Elvis \\
        Harvard-Smithsonian Center for Astrophysics, 60 Garden St, Cambridge, MA 02138}
\date{November 4, 2009}

\begin{document}

\maketitle

\label{firstpage}

\begin{abstract}
We use 62,185 quasars from the Sloan Digital Sky Survey DR5 sample and standard virial mass scaling laws based on the widths of H$\beta$, Mg{\small II}, and C{\small IV} lines and adjacent continuum luminosities to explore the maximum mass of quasars as a function of redshift, which we find to be sharp and evolving.  This evolution is in the sense that high-mass black holes cease their luminous accretion at higher redshift than lower-mass black holes.  Further, turnoff for quasars at any given mass is more highly synchronized than would be expected given the dynamics of their host galaxies.  We investigate potential signatures of the quasar turnoff mechanism, including a dearth of high-mass quasars at low Eddington ratio.  These new results allow a closer examination of several common assumptions used in modeling quasar accretion and turnoff.
\end{abstract}


\section{Introduction}
\label{sec:intro}

Supermassive black holes (SMBH) are found today at the centers of nearly every galaxy where there have been sensitive searches.  These SMBH are the remnants of bright quasars which are found in many galaxies at redshifts $z \sim 2$.  The mechanisms involved in the quasar {\em turnoff} process are not well understood (cf. Thacker et al. 2006\nocite{turnoffreview}).  In this work, we combine the large quasar sample provided by the Sloan Digital Sky Survey (SDSS) \cite{Schneider2007} with recent advances in virial mass estimation \cite{Vestergaard2006,Shen2008} to develop new constraints on quasar turnoff.

It had been previously thought that most quasars were within a factor of $\sim 10$ of their Eddington luminosity \cite{Kollmeier2006}, and quasars have been commonly modeled as `light-bulbs', operating either near Eddington or without substantial luminous accretion, although this is likely an oversimplification \cite{Hopkins2008}.  In Paper I \cite{Steinhardt2009}, we used the quasar distribution in the mass-luminosity plane as a function of redshift.  Had the relationship between quasar mass and luminosity been trivial (as implied by the the light-bulb approximation) the mass-luminosity plane would not present an improvement over the quasar mass function or luminosity function.  Instead, we found a more complex relationship between mass and luminosity than had been previously assumed.  Several new puzzles have emerged from this view of the quasar distribution, including a sub-Eddington boundary that restricts luminosities of the highest-mass quasars at every redshift to lie below their Eddington limit.  We also noted an upper mass limit varying with redshift.  In this work, we use this new, multi-dimensional view of the quasar distribution to investigate this high-mass `turnoff' in detail.

We use these results to evaluate three basic assumptions: (1) that the dynamics of quasar turnoff are closely linked to the dynamics of the host galaxy, (2) that a segregation of quasars by mass is equivalent to a segregation of quasars by luminosity, (a corollary of the light-bulb approximation), and (3) that SMBH undergo no fundamental change during turnoff, but rather run out of fuel (which implies that an individual SMBH might have many luminous episodes during its lifetime and would undergo another one today given sufficient fuel).  

%In previous work \cite{Steinhardt2009a}, we showed what the M-L plane says about accretion.  Now, we're going to show that slicing it in different ways is interesting as well.

In \S~{\ref{sec:revml}}, we briefly review the SDSS sample and exhibit its behavior in the $M-L$ plane.  A full discussion of this sample and its uncertainties can be found in Schneider et al. (2003) and Paper I\nocite{Schneider2003,Steinhardt2009}.  In \S~\ref{sec:permanent}, we establish that massive quasars are indeed turning off, and doing so earlier than lower-mass quasars.  In \S~\ref{sec:signatures}, we use this to select individual quasars near turnoff and search for signatures of their turnoff mechanisms.  In \S~\ref{sec:synch}, we show that the synchronization of turnoff of different quasars at a given mass is closer than that of the dynamics of their host galaxies.  We also show that the synchronization of quasars at fixed mass is characteristically different from that of quasars at fixed luminosity as reported in Amarie et al. (2009)\nocite{Amarie2009}.  In \S~\ref{sec:discussion}, we evaluate the three basic assumptions above in light of this new evidence.

\section{SDSS Quasars in the M-L Plane}
\label{sec:revml}

Black hole masses for 62,185 of the 77,429 SDSS DR5 quasars \cite{Schneider2007} were determined by Shen et al. (2008) using H$\beta$- and C{\small IV}-based virial mass estimates from Vestergaard \& Peterson (2006)\nocite{Vestergaard2006} and Mg{\small II}-based estimators from McLure \& Dunlop (2004)\nocite{McLure2004}.  As discussed elsewhere \cite{Shen2008,Risaliti2009,Steinhardt2009}, C{\small IV}-based estimation is substantially less accurate than H$\beta$ or Mg{\small II}, while Mg{\small II} has a smaller statistical uncertainty than H$\beta$.  Vestergaard \& Peterson (2006) report an uncertainty for virial mass estimation of $\sim 0.4$ dex from calibration against reverberation mapping, while in Paper III we discuss the possibility that the statistical uncertainty may be closer to $\sim 0.15$ dex and the remaining deviations due to systematic effects.

\begin{figure*}
  \epsfxsize=7in\epsfbox{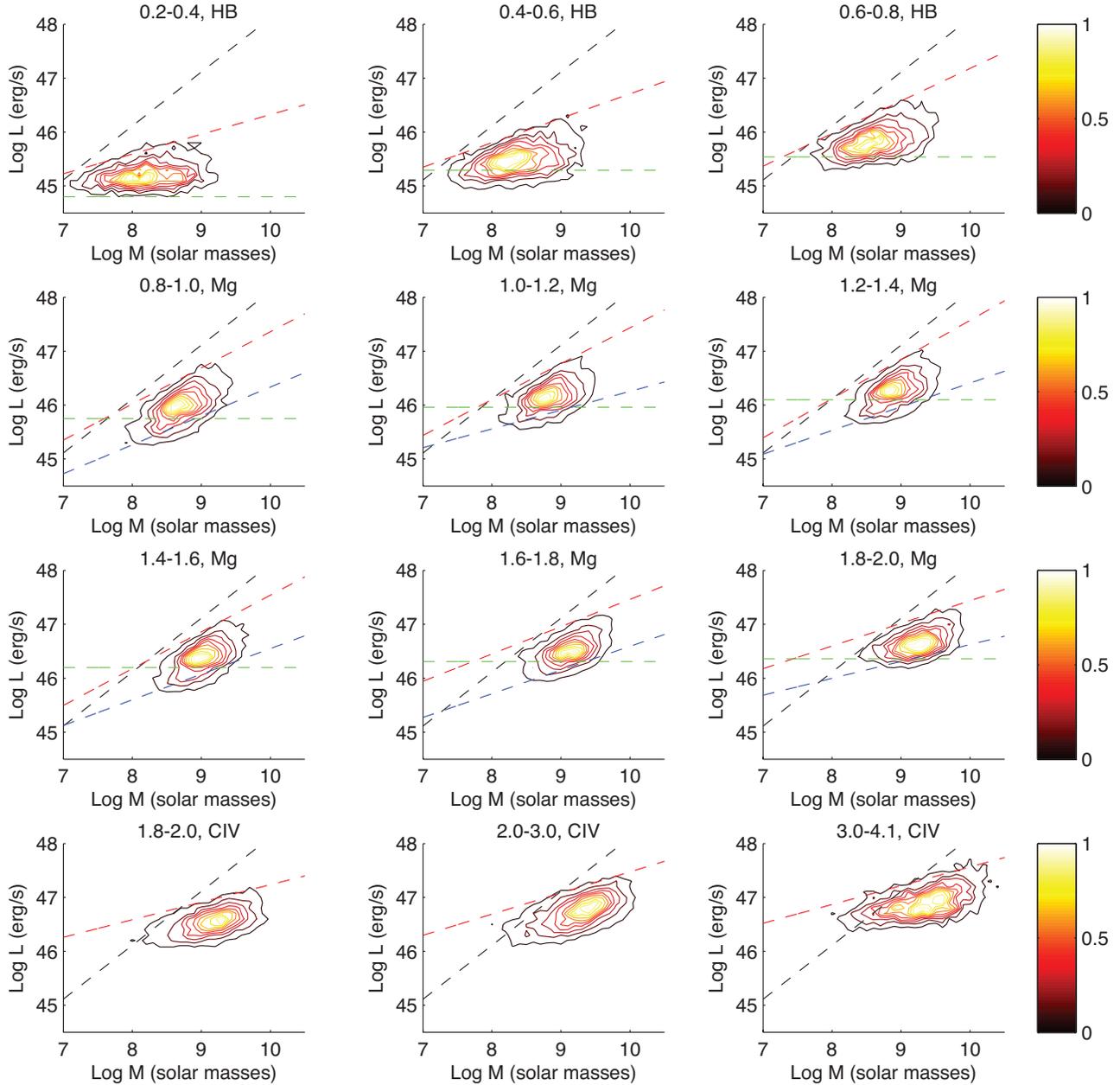}
\caption{A contour plot of the black hole mass-bolometric luminosity distribution in 12 different redshift ranges, as described in Steinhardt \& Elvis (2009).  In each redshift bin, the quasar number density has been normalized to the peak number density.  The top three panels are from the low-redshift sample, i.e., with H$\beta$-based mass estimates; the middle six panels are from the medium-redshift sample, i.e., with Mg{\small II}-based mass estimates; and the bottom three panels are from the high-redshift sample, i.e., with C{\small IV}-based mass estimates.  The red dashed line is drawn at the sub-Eddington boundary (Paper I) and the blue dashed line in the Mg{\small II} panels are drawn at the high-mass, low-luminosity boundary.  The green dashed line is drawn at approximately 19.1 in $i$ band for  the average redshift in each panel; below this line objects are from the serendipitous sample.}
\label{fig:allzcontour}
\end{figure*}

As in Paper I, we divided the sample into redshift bins of size $0.2$ in $z$ for $z < 2.0$ and larger redshift bins for $z > 2.0$.  Table {\ref{table:bins}} contains summary statistics on the objects in each bin.  Within each bin, there is a distribution of black hole masses spanning $\sim 2$ dex (Figure \ref{fig:allzcontour}).   
\begin{table}
\caption{Summary statistics on quasars in the 13 redshift and emission line bins.}
\begin{tabular}{|c|c|c|c|c|c|}
\hline 
ID & $z$ & $N$ & $<\log L> (erg/s)$ & $<\log M/M_\odot>$ & Line \\
\hline 
1 & 0.2-0.4 & 2690 & 45.25 & 8.27 & H$\beta$ \\
2 & 0.4-0.6 & 4250 & 45.54 & 8.44 & H$\beta$ \\
3 & 0.6-0.8 & 3665 & 45.89 & 8.69 & H$\beta$ \\
4 & 0.6-0.8 & 4727 & 45.80 & 8.59 & Mg{\small II} \\
5 & 0.8-1.0 & 5197 & 46.02 & 8.76 & Mg{\small II} \\
6 & 1.0-1.2 & 6054 & 46.21 & 8.89 & Mg{\small II} \\
7 & 1.2-1.4 & 7005 & 46.32 & 8.96 & Mg{\small II} \\
8 & 1.4-1.6 & 7513 & 46.43 & 9.07 & Mg{\small II} \\
9 & 1.6-1.8 & 6639 & 46.57 & 9.18 & Mg{\small II} \\
10 & 1.8-2.0 & 4900 & 46.71 & 9.29 & Mg{\small II} \\
11 & 1.8-2.0 & 4627 & 46.60 & 9.20 & C{\small IV} \\
12 & 2.0-3.0 & 7079 & 46.79 & 9.30 & C{\small IV} \\
13 & 3.0-4.1 & 2859 & 46.98 & 9.38 & C{\small IV} \\
\hline  
\end{tabular}
\label{table:bins}
\end{table}

\begin{figure}
  \epsfxsize=3in\epsfbox{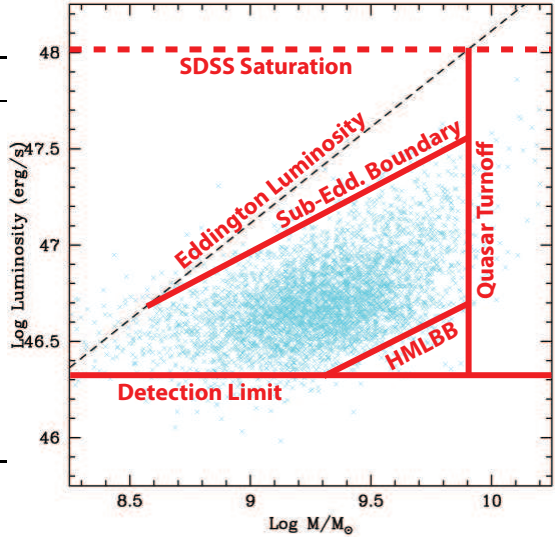}
\caption{The SDSS quasar locus of the $H\beta$ mass sample in the $M-L$ plane at redshift $0.2 < z < 0.4$.  The locus should be bounded by SDSS {\bf detection limits}, $L_{Edd}$ (dashed line), and on the high-mass end by an unknown mechanism responsible for {\bf quasar turnoff}.  There is an additional {\bf sub-Eddington boundary} (Steinhardt \& Elvis 2009) with slope below that of $L_{Edd}$.  Finally, high-mass, low-luminosity quasars are sparse, suggesting the possible presence of an additional boundary ({\bf HMLLB}) or added complexity in the nature of quasar turnoff.  The bright-object {\bf SDSS saturation} limit does not intersect the quasar locus.  These lines are defined quantitatively either here or in Paper I.}
\label{fig:hb1}
\end{figure}

 Figure \ref{fig:allzcontour} shows quasar loci in the $L-M$ plane for each of the 12 redshift bins as contour plots.  An individual quasar locus in Figure \ref{fig:hb1} has been labeled at its various boundaries.  The quasar locus at each redshift is bounded by SDSS detection limitations (bottom), the Eddington luminosity (dashed line, upper-left), a sub-Eddington boundary (top) discussed at length in Steinhardt \& Elvis (2009)\nocite{Steinhardt2009}, a high-mass cutoff (right) associated with quasar turnoff and discussed in \S~\ref{sec:permanent}, and possibly an additional high-mass, low-luminosity boundary discussed in \S~\ref{sec:signatures}.

No quasars in the catalog have luminosity greater than $L_{Edd}$ with statistical significance, while a sub-Eddington boundary is present in each panel.  In each panel, the low-luminosity boundary on the quasar locus is the only boundary due to SDSS detection limitations.  In particular, SDSS would be capable of detecting quasars past the high-mass and high-luminosity boundaries of the quasar locus in each panel and including them in the catalog if they were to exist.  

\section{Massive Quasars Are Turning Off}
\label{sec:permanent}

As shown in Table \ref{table:bins}, both the maximum luminosity and the maximum mass of the quasar locus are monotonically decreasing towards lower redshift.  Figure \ref{fig:allzcontour} hints the same might be true of the average luminosities and masses of the quasar populations in each bin.  While it has been observed that individual quasars have variable luminosity \cite{Mathews1963}, we know of no mechanism by which the central black hole can substantially reduce its mass.  Therefore, we must interpret Figure \ref{fig:allzcontour} as showing us that the most massive quasars in any cosmological epoch are in the midst of disappearing.  

Each redshift bin in \ref{table:bins} is of an equal size in redshift, and therefore a different size in comoving volume.  It is therefore expected that the most massive quasars become rarer at low redshift; Mg{\small II}-based mass estimates at $1.8 < z < 2.0$ encompass a comoving volume 15 times larger than Mg{\small II}-based mass estimates at $0.6 < z < 0.8$.  If the number density of massive quasars were constant, there should be 15 times as many high-mass quasars visible at $1.8 < z < 2.0$ than at $0.6 < z < 0.8$.  

\begin{figure}
\leavevmode
      \epsfxsize=3in\epsfbox{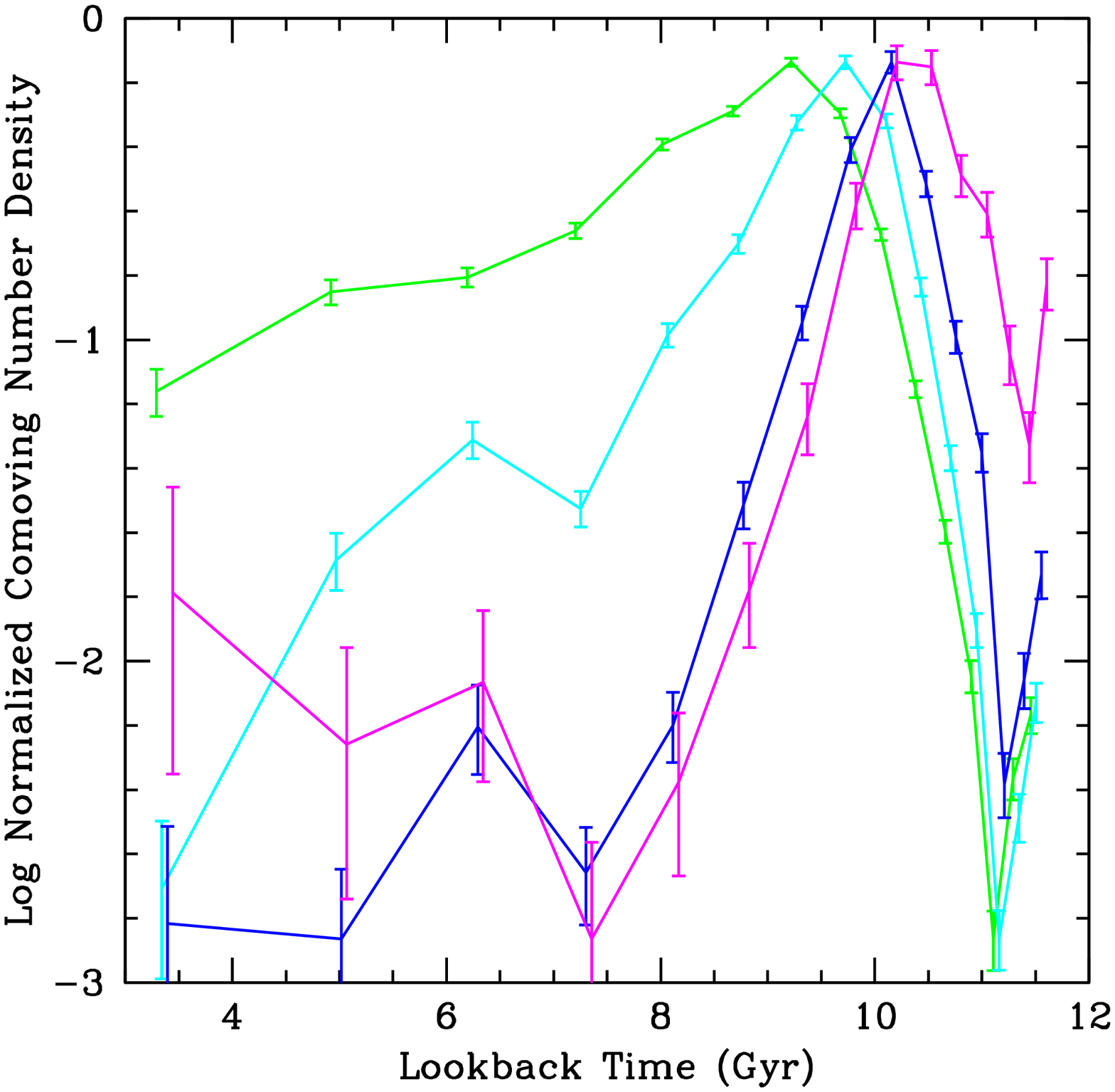}
  \caption{Raw comoving number density evolution of quasars in four different mass bins at $0.2 < z < 3.2$, normalized to peak.  Poisson uncertainties are indicated and the comoving number density at each mass is normalized to its peak value. Each curve contains quasars in a 0.25 dex range in mass: green $9.0 < \log M/M_{\odot} < 9.25$, cyan 9.25-9.5, blue 9.5-9.75, and magenta 9.75-10.0.}
\label{fig:turnoffall}
\end{figure}

The number density at a given mass as a function of redshift for all quasars in the SDSS catalog is shown in Figure \ref{fig:turnoffall}.  In all but the lowest mass bins, the catalog comoving number density rises to a peak between a redshift of 1 and 2, then declines towards lower redshift.
However, the luminosity distribution for quasars at fixed mass and redshift is cut off on the low-$L$ end by the SDSS detection limit for many masses (e.g., at $9.0 < \log M/M_{\odot} < 9.25$ in Figure \ref{fig:turnoffall}). 
\begin{figure}
\leavevmode
      \epsfxsize=3in\epsfbox{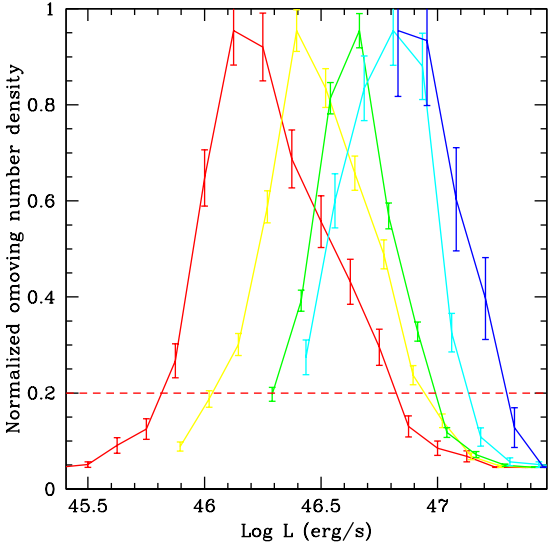}
  \caption{Luminosity distribution of SDSS quasars with masses $9.0 < \log M/M_{\odot} < 9.25$ in different redshift bins:  red $0.8 < z < 1.0$, yellow $1.2 < z < 1.4$, green $1.6 < z < 1.8$, cyan $2.0 < z < 2.2$, and blue $3.0 < z < 3.2$.  Poisson uncertainties are indicated and at each redshift the number density is normalized to its peak value.  The red, yellow, and green curves dip below 20\% of peak at the detection limit and therefore are considered sufficiently complete, while the cyan and blue curves are not.}
\label{fig:lumdist}
\end{figure}
Therefore, the total number density of quasars at a given mass is only known in some redshift bins, for which only a negligible fraction of quasars at that mass lie below the SDSS limit.  To address this bias, we only include bins for which the number density at the detection limit is $< 20\%$ of the peak.  Figure \ref{fig:lumdist} shows a sample cut for quasars with $9.0 < \log M/M_\odot < 9.25$.  We include several bins where there are fewer than ten total quasars, but none at the detection threshold.  Restricting Figure \ref{fig:turnoffall} to those points produces the number density-lookback time plot in Figure \ref{fig:turnoffcut}. 
\begin{figure}
\leavevmode
      \epsfxsize=3in\epsfbox{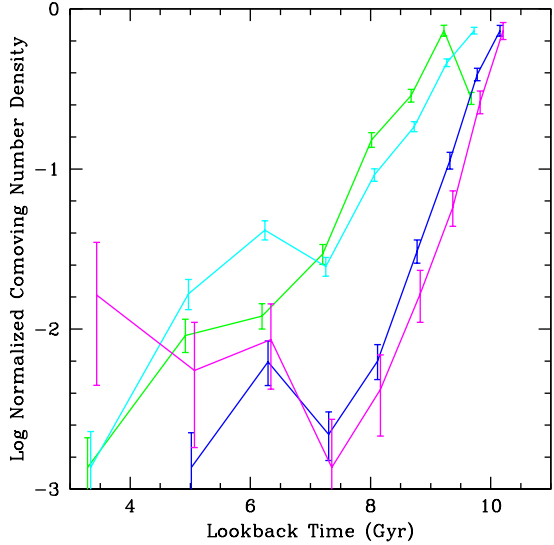}
  \caption{Completeness-selected comoving number density evolution with lookback time for quasars in different mass bins at $0.2 < z < 2.0$.  Displayed data is restricted to bins in which nearly all quasars could be detected and included in the SDSS catalog.  Poisson uncertainties are indicated and the comoving number density at each mass is normalized to its peak value.  Each curve contains quasars in a 0.25 dex range in mass: green $9.0 < \log M/M_{\odot} < 9.25$, cyan 9.25-9.5, blue 9.5-9.75, and magenta 9.75-10.0.}
\label{fig:turnoffcut}
\end{figure}  

\begin{figure}
\leavevmode
      \epsfxsize=3in\epsfbox{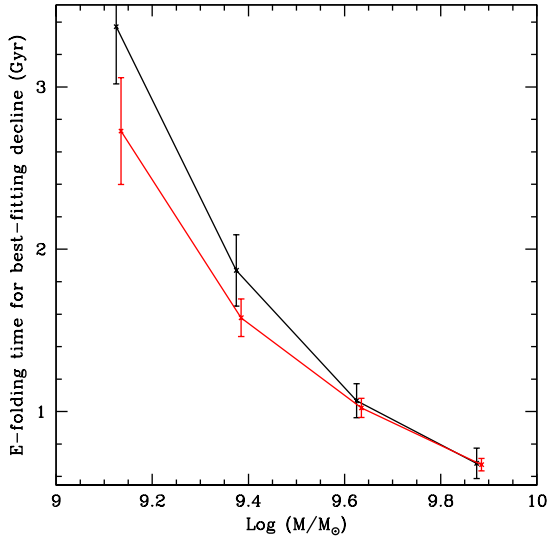}
  \caption{Timescale for best-fitting exponential decay in comoving number density of quasars as a function of mass.  The black points have been fit using both H$\beta$ and Mg{\small II} masses, while the red points only use Mg{\small II} mass estimates.}
\label{fig:turnoffslope}
\end{figure}

The number density distributions in Figure \ref{fig:turnoffcut} clearly decline with increasing cosmic time (decreasing redshift).  The higher-mass curve appears to decline faster.  To investigate this possibility, we fit this decline as an exponential decay at each mass.  The results are summarized in Table \ref{table:decays} and Figure \ref{fig:turnoffslope}.  
\begin{table}
\caption{Best-fitting exponential decays for declining comoving number densities of quasars in different mass bins.  Also included are the best-fitting declines using only Mg{\small II}-based mass estimation.}
\begin{tabular}{|c|c|c|c|c|}
\hline 
$\log M/M_\odot$& $e$-folding (Gyr) & $\chi^2_\nu$ & Mg{\small II} only (Gyr) & $\chi^2_\nu$ \\
\hline 
9.0--9.25 & $3.37 \pm 0.35$ & 9.37 & $2.73 \pm 0.33$ & 4.05 \\
9.25--9.5 & $1.87 \pm 0.22$ & 17.83 & $1.58 \pm 0.12$ & 2.67 \\
9.5--9.75 & $1.07 \pm 0.11$ & 8.89 & $1.02 \pm 0.06$ & 1.92 \\
9.75--10.0 & $0.68 \pm 0.09$ & 2.67 & $0.67 \pm 0.04$ & 0.65 \\ 
\hline  
\end{tabular}
\label{table:decays}
\end{table}
At higher masses, characteristic time of the decline in comoving number density is indeed shorter.  The higher mass uncertainty in the H$\beta$-based mass estimates leads to a better fit when only the Mg{\small II}-based mass estimates are used, but both estimates are in agreement at each mass.  The high $\chi^2$/DOF values either suggest that the decay cannot be well-modeled as exponential or that Poisson errors are underestimates of the true uncertainty in the comoving number density.  The latter almost certainly applies, as the mass uncertainty is $\sim 0.4$ dex \cite{Vestergaard2006} while bins are just $0.25$ dex in width (although in Paper III we discuss evidence that the statistical uncertainty in virial mass estimation may be smaller than previously believed).  However, the decline may not be exponential.  An improvement in the detection limit would allow a measurement of the comoving number densities over a wider range of redshift, and therefore might allow a better determination of the exact nature of their declines.

The timescales for decline $\tau_d$ are consistent with a dropping exponential with mass (Figure \ref{fig:turnoffslope}).  Using all mass estimates available, $\tau_d \propto M^{-2.097 \pm 0.004}$ with a $\chi^2$/DOF of 0.60, while the timescales using only Mg{\small II} mass estimates are proportional to $\tau_d \propto M^{-1.809 \pm 0.155}$ with a $\chi^2$/DOF of 0.626.  The fit uncertainty using all mass estimates is small, but the uncertainties in individual measurements of $\tau_d$ indicate that the true uncertainty in slope is considerably higher than the fit uncertainty.

The comoving number density of high-mass quasars (Figure \ref{fig:turnoffcut}) shows that high-mass quasars are indeed less common at lower redshift, with the highest-mass quasars having the most rapid decline in number density.  At each redshift, the SDSS catalog contains a snapshot of a population of the highest-mass quasars set to decline.  Massive quasars indeed turn off, with the highest-mass quasars turning off first.  In \S~{\ref{sec:signatures}}, we search for signatures of this quasar ``downsizing' in an effort to study the causes and mechanisms involved.

\section{Signatures of Quasar Turnoff}
\label{sec:signatures}

At every redshift, the high-mass quasar population as a whole is declining rapidly as in \S~\ref{sec:permanent}.  It follows that a snapshot of that population must catch many individual high-mass quasars that will shortly undergo turnoff.  Further, if the turnoff process is gradual, it is very likely that the SDSS catalog will contain quasars in the midst of turnoff.  For example, if quasar turnoff takes a minimum of 40 Myr as suggested by Hopkins et al. (2008)\nocite{Hopkins2008}, we should expect $\sim 200$ quasars with $M_{BH} > 9.5$ at $1.8 < z < 2.0$ to turnoff in the next 1 Gyr, and therefore $\gtrsim 8$ to be in the midst of turnoff while emitting the light we observe.  

Here, we search for anomalous properties of the highest-mass quasars in each redshift bin the hopes that they will provide signatures of the mechanism by which quasars turn off.

%\subsection{Line Amplitudes}
%
%The ratio of line fluxes might be a poor probe of the quasar accretion disk, since the continuum spectral energy distribution is not constant.  In addition to the line flux, we can also consider the line amplitude as judged by the line equivalent width.  The ratio of 

\subsection{The High-Mass, Low-Luminosity Boundary}

We can attempt to probe the turnoff rate by looking at the luminosities of the most massive quasars.  Consider one of the most massive quasars in a given redshift bin that lies at a luminosity near the sub-Eddington boundary (SEB, Paper I).  As discussed above, this is a snapshot of a quasar which will shortly begin its turnoff.  If this turnoff is gradual, a snapshot of this quasar at a later time must show a mass larger than any quasar at the SEB but at a luminosity well below the SEB.  Since $M_{BH}$ for the most massive quasars declines with decreasing redshift, a more gradual turnoff process will result in a larger population of massive,low-Eddington ratio quasars with mass more greatly exceeding the most massive quasars lying on the SEB.  

\begin{figure}
  \epsfxsize=3in\epsfbox{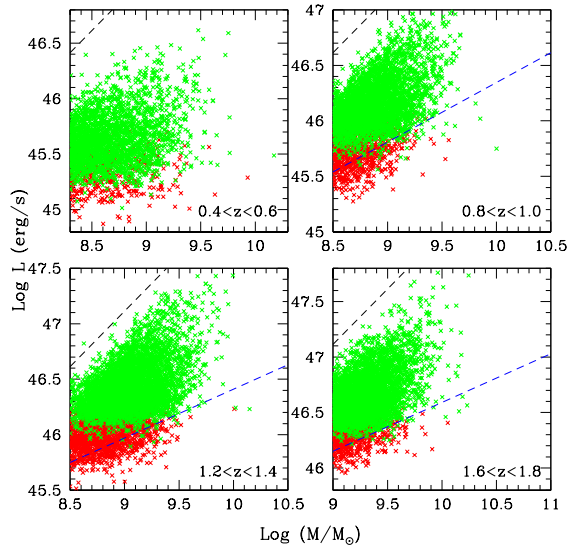}
\caption{The high-mass, low-luminosity end of the quasar mass-luminosity plane in four redshift bins.  Quasars at $0.4 < z < 0.6$ are displayed with H$\beta$ mass estimates, while quasars at $0.8 < z < 1.0$, $1.2 < z < 1.4$, and $1.6 < z < 1.8$ use Mg{\small II} mass estimates.  Quasars brighter than $i = 19.2$ are shown in green, while quasars in the serendipitous sample are shown in red.  High-mass, low-luminosity quasars are scarce in three panels, indicating the presence of an HMLLB (blue).}
\label{fig:highmldist}
\end{figure}

However, Figure \ref{fig:allzcontour} shows that not only is there no substantial population in any redshift bin $z > 0.6$ of low-luminosity quasars with masses higher than the rest of the quasar population, but also that the highest mass quasars at every redshift are restricted to luminosities well above the SDSS detection threshold.  The luminosity distribution of high-mass quasars in four representative redshift bins is shown in Figure \ref{fig:highmldist}.  

At almost every mass and redshift the peak quasar number density occurs at a luminosity above the SDSS detection threshold (Figure \ref{fig:allzcontour}).  In Paper I, we defined the SEB by looking at the decline from that peak on the high-luminosity end.  Here we similarly define a boundary on the low-luminosity end as the point at which the number density has dropped to 20\% of peak.  Note that because the peak number density is not constant across mass bins, this boundary is different from a 20\% contour on the entire locus.  Contours of quasar number density in the $M-L$ plane also exclude a high-mass, low-luminosity region (Figure \ref{fig:allzcontour}).

We consider a high-mass, low-luminosity boundary (HMLLB) to be present when for at least the highest 0.5 dex in mass, the boundary lies significantly above the SDSS detection threshold.  Best-fitting lines to the HMLLB, where present, are summarized in Table \ref{table:hmllb}.
\begin{table}
\caption{Best-fitting linear HMLLB slopes at $0.2 < z < 2.0$.  There is no HMLLB above the SDSS detection threshold at $z < 0.6$.  Where an HMLLB exists, its slope seems not to be a constant fraction of $L_{Edd}$, but rather at or below that of the SEB at the same redshift.  The range of $L/L_{Edd}$ from the SEB to the HMLLB at the high-mass end is also included.}
\begin{tabular}{|c|c|c|c|c|}
\hline 
$z$ & Line & Slope & SEB slope & $L/L_{Edd}$ range (\%) \\
\hline 
0.2-0.4 & H$\beta$ & none & $0.37 \pm 0.02$ & 1-8 \\
0.4-0.6 & H$\beta$ & none & $0.45 \pm 0.03$ & 1-11 \\
0.6-0.8 & H$\beta$ & $0.60 \pm 0.19$ & $0.60 \pm 0.06$ & 1-20 \\
0.6-0.8 & Mg{\small II} & $0.67 \pm 0.10$ & $0.61 \pm 0.10$ & 3-32 \\
0.8-1.0 & Mg{\small II} & $0.54 \pm 0.03$ & $0.67 \pm 0.09$ & 3-25 \\
1.0-1.2 & Mg{\small II} & $0.35 \pm 0.03$ & $0.67 \pm 0.05$ & 2-25 \\
1.2-1.4 & Mg{\small II} & $0.44 \pm 0.07$ & $0.73 \pm 0.05$ & 2-31 \\
1.4-1.6 & Mg{\small II} & $0.47 \pm 0.02$ & $0.68 \pm 0.08$ & 3-33 \\
1.6-1.8 & Mg{\small II} & $0.44 \pm 0.04$ & $0.50 \pm 0.10$ & 3-25 \\
1.8-2.0 & Mg{\small II} & $0.31 \pm 0.03$ & $0.42 \pm 0.06$ & 3-24 \\
\hline  
\end{tabular}
\label{table:hmllb}
\end{table}
Poisson uncertainties on the number density of objects at the boundary suggest a relative uncertainty between the boundary at different masses.  We use these relative uncertainties in the determination of a best-fitting line.  At each sufficiently high redshift ($z > 0.6$), there is a HMLLB present.  Below this redshift, the peak number density lies near the SDSS detection limit, so it is possible that a HMLLB would exist with improved detection.  The HMLLB is not at a constant fraction of $L_{Edd}$ (slope of 1.0), but rather at a slope at or below that of the sub-Eddington boundary.  

At $0.6 < z < 0.8$, where there are both Mg{\small II} and H$\beta$ masses available, the HMLLB have similar slopes but start at a $\sim 0.2$ dex lower mass in the Mg{\small II} quasar locus than in the H$\beta$-based locus.  Onken \& Kollmeier (2008)\nocite{Onken2008} suggest that Mg{\small II} masses may need to be recalibrated against H$\beta$ masses, and Risaliti, Young, \& Elvis (2009)\nocite{Risaliti2009} compare H$\beta$ and Mg{\small II} masses to suggest $\log[M_{BH}(H\beta)]=1.8\times\log[M_{BH}(Mg{\small II})]-6.8.$.  Applying this correction would produce a Mg{\small II}-based HMLLB slope of $0.37 \pm 0.06$ at $0.6 < z < 0.8$.  Both this corrected slope and the uncorrected HMLLB slope of $0.67 \pm 0.10$ are consistent with the H$\beta$-based slope of $0.60 \pm 0.19$.  While the SEB at this redshift implies the Mg{\small II} mass correction may not be required for the most massive and most luminous objects at each redshift, the HMLLB only provides an inconclusive test of these corrections at high mass and low luminosity.

Possible explanations for the dearth of high-mass, low-luminosity quasars can be divided into three categories:
\begin{enumerate}
\item{The SDSS pipeline might not select such objects as quasars, either due to an error in selection or because there is a high-mass, low-luminosity boundary (HMLLB) at which quasars gradually transition to an object classified differently (such as a Seyfert galaxy) but with a similar spectral energy distribution to a quasar.}
\item{Their masses might be systematically underestimated or luminosities systematically overestimated.}
\item{There might be a high-mass, low-luminosity boundary at which quasars very rapidly transition to an object with a very different spectral energy distribution outside the SDSS locus, spending little time as a transitional object similar to a quasar.}
\end{enumerate}
We will consider the first two possibilities below and the third in \S~\ref{sec:discussion}.
%In Figure \ref{fig:11}, we show the high-mass tail of the quasar locus in four redshift bins in an effort to search for this effect.

\begin{figure}
  \epsfxsize=3in\epsfbox{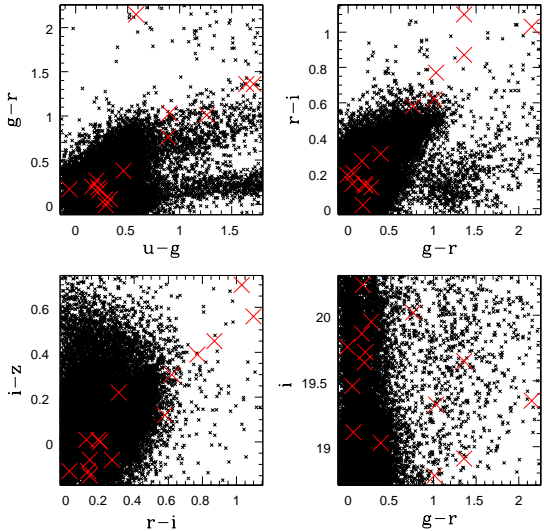}
\caption{Colors of the 15 quasars (red) at $1.4 < z < 1.6$ with $M > 10^{9.5} M_\odot$ and $L < 0.025 L_{Edd}$ compared to all SDSS quasars (black).  These high-mass, low-luminosity quasars do not all lie along a common photometric selection boundary but are comingled with other SDSS quasars.}
\label{fig:turnsel}
\end{figure}

SDSS selection for well-measured objects above the detection limit consists of identifying the object from its magnitudes in the five SDSS spectral bands.  If there is an HMLLB at which SDSS mistakenly ceases to select objects as quasars, this would imply that quasars at the HMLLB are also at a boundary of the SDSS quasar locus.  The sharpest HMLLB (with the fewest points lying below the HMLLB and the best-determined slope) occurs in the $1.4 < z < 1.6$ redshift bin, so it is the best candidate for a selection effect.  At that redshift, there are 15 quasars with $M > 10^{9.5} M_\odot$ and $L/L_{Edd} < 0.025$.  In Figure \ref{fig:turnsel}, the colors of these 15 objects are shown in four slices of the five-dimensional color space used by SDSS in its photometric selection.  
Their locations are not tightly clustered in color space, although a larger fraction lie at the outer edges of the SDSS loci than would be expected if they were drawn randomly from the SDSS quasar catalog.  However, the HMLLB does not lie along a photometric selection boundary, nor are there additional selection criteria that these objects are close to \cite{Richards2002}.  SDSS selection is not responsible for the HMLLB.

While a mass-dependent systematic error in luminosity determination seems plausible, masses at the HMLLB at lower redshift are at the low-mass end of the quasar sample at higher redshift.  As discussed in Paper I\nocite{Steinhardt2009}, when quasars determined to have identical masses are at the low-mass end of the quasar sample, they reach $L_{Edd}$ but do not surpass it.  Therefore, any systematic over-estimation of quasar luminosity of the sort required to explain the HMLLB either also requires that no high-redshift quasar be near $L_{Edd}$ or that the HMLLB is caused by a coincidence between redshift-dependent and mass-dependent systematic errors that almost entirely cancel for quasars at lower masses in every redshift bin while producing the HMLLB at higher masses.  Such fine-tuning seems highly implausible, although it cannot be ruled out from direct observational evidence.

A systematic error in mass determination at the HMLLB requires virial masses to be incorrect for faint quasars with very broad emission lines.   For Mg{\small II} in particular, the wings of wide emission lines with low equivalent width might be difficult to discern from slightly redder Fe{\small II} lines.  To produce the HMLLB, at $1.4 < z < 1.6$ high-mass and low-luminosity objects would need to have their masses underestimated by $\sim 0.5$ dex, if the highest-mass objects have identical mass at every luminosity (Figure \ref{fig:allzcontour}).  If turnoff is slow and the highest mass objects are more massive at low $L$ then a greater factor would be required.  Because virial masses depend upon the square of the H$\beta$, Mg{\small II}, or C{\small IV} full-width half-maximum (FWHM), an error of $\sim 0.5$ dex in mass would require an error of $\sim 0.25$ dex in FWHM.  
\begin{figure}
  \epsfxsize=3in\epsfbox{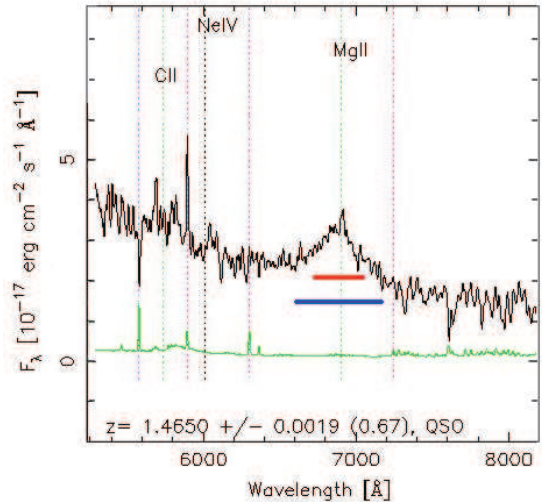}
\caption{Spectrum of SDSS J171737.29+535647.3, a quasar lying on the HMLLB at $z=1.46.$  The FWHM of Mg{\small II} as determined by Shen et al. (2008) is indicated in red.  The FWHM required to increase the virial mass estimate by 0.5 dex is indicated in blue.}
\label{fig:mgfwhm}
\end{figure}

Figure \ref{fig:mgfwhm} shows the spectrum of a quasar lying near the $1.4 < z < 1.6$ HMLLB.  This spectrum has the median signal-to-noise ratio for the $1.4 < z < 1.6$ HMLLB of 3.36, while quasars on the HMLLB range in signal-to-noise from 2.26 to 6.29.  The FWHM measured by the SDSS pipeline is shown in red.  To increase the virial mass estimate by 0.5 dex, the FWHM would need to instead be of the size indicated by the blue line, which is clearly too broad and is closer to the entire width of the line itself than to its FWHM.  
%Higher quality spectra would be even less likely to be off by the $\sim 80$\% or more in FWHM determination required to remove the HMLLB.

There is an window of redshift at $0.6 < z < 0.8$ in which both H$\beta$-based and Mg{\small II}-based mass estimates are available.  If the Mg{\small II}-based estimates are systematically too small near the HMLLB due to Fe contamination, the H$\beta$-based estimate for the same objects should be systematically larger.  For all 3505 objects at $0.6 < z < 0.8$, the mean $M_{Mg}-M_{H\beta} = -0.06$ dex, with a standard deviation of $0.28$ dex.  Of the 56 objects closest to the HMLLB (as determined by Mg{\small II} masses), the mean $M_{Mg}-M_{H\beta} = 0.05$ dex smaller than those of Mg{\small II}, with a standard deviation of $0.44$ dex.  Because these are at the high-mass tail of the Mg{\small II} mass distribution, they are more likely to be objects where the Mg{\small II} mass is larger than the H$\beta$ mass if their statistical uncertainties are uncorrelated.  So, the atypically large mean $M_{Mg}-M_{H\beta}$ for this sample should be expected.  However, neither the entire population nor only quasars near the HMLLB have a difference close to the the $\sim 0.5$ dex correction required to remove the HMLLB.

Finally, since virial mass estimation is calibrated from just $\sim 30$ objects \cite{Vestergaard2006}, the relation may be poorly calibrated for unusual objects, perhaps including those near the HMLLB.  In particular, it is possible that the line FWHM and continuum are measured correctly but virial masses are systematically low near the HMLLB.  This possibility cannot be ruled out directly from available data.  Virial mass estimates can be very nearly derived theoretically under the two assumptions that (1) peculiar velocities in the broad-line region are nearly virial and (2) the luminosity is nearly $L \propto R^2$.  The successful calibration of virial mass estimates against reverberation mapping suggests that these two assumptions hold for most quasars.  Breakdown in virial mass estimation near the HMLLB would be an indicator that the properties of quasars change near the HMLLB.

In conclusion, the HMLLB is not caused by (1) erroneous SDSS selection or (2) errors in determining spectral parameters used in virial mass estimation.  Incorrect luminosity determination near the HMLLB would require a fine-tuning between redshift-dependent and mass-dependent systematic errors.  The most likely explanation for the HMLLB appears to be a transition in the properties of the quasar, which may be manifested either as a rapid change in the quasar SEB (leading to an object that no longer resembles a quasar in the five SDSS color bands) or as a breakdown in virial mass estimation near the HMLLB (e.g., if the broad line region is no longer predominantly virial).

\subsection{Line Ratios}
Different emission lines come from transitions with different ionization potentials, and are interpreted as being emitted at different radii from the central black hole\cite{Petersonbook}.  In particular, Mg{\small II} is emitted furthest from the central black hole and C{\small IV} nearest.  Changes in quasar structure near turnoff, either out in the broad-line region (BLR) or closer in at the optical/UV emitting part of the accretion disk, may show up as unusual line ratios.  The redshift ranges where we can see either the Mg{\small II} and H$\beta$ or Mg{\small II} and C{\small IV} lines in the same spectrum allow us to test this idea.  In Figure 10 we show these two line ratios as a function of mass. 
\begin{figure}
  \epsfxsize=3in\epsfbox{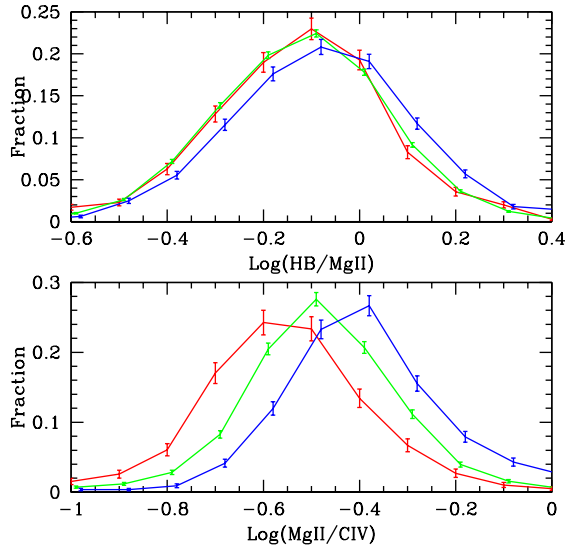}
\caption{Distribution in three mass bins of the flux ratio of the H$\beta$ to Mg{\small II} (top) and Mg{\small II} to C{\small IV} (bottom) emission lines at redshifts where both lines are visible in SDSS spectra.  At top ($z \sim 0.6$), the boundaries between bins are $\log M/M_\odot = $8.0 and 9.0, while on the right ($z \sim 2.0$), the mass boundaries are at $\log M/M_\odot = $ 9.0 and 9.5.  The high-mass distribution is drawn in blue while the low-mass distribution is drawn in red in each panel.  The indicated uncertainties are Poisson errors and are likely underestimates.}
\label{fig:10}
\end{figure}

For $0.6 < z < 0.8$, the H$\beta$/Mg{/small II} ratio has a similar distribution at all black hole masses, although the very highest-mass objects (blue, Figure \ref{fig:10}) might hint at a slight deviation (3\% chance to be identical, A Kolmogorov-Smirnov test).  This similarity suggests that, high-mass and low-mass quasars at $z \sim 0.6$ have both similar BLRs and ionizing continuum shapes.  There is no difference in either the virial mass or bolometric luminosity estimation techniques between high-mass and low-mass quasars, and this similarity is at least partial confirmation that this might be reasonable.

A majority of the 723 quasars at $M > 9.4$ will turn off by $z=0.4$, i.e., within their next $\sim 1.4 $Gyr.  Since we see no difference in the H$\beta$/Mg{\small II}   line ratio, this either means that the process by which they cease luminous accretion does not substantially change this ratio or that the turnoff is sufficiently rapid that we see no signature of it in Figure \ref{fig:10}.  Since there are $\sim 400$ quasars that turn off within $\sim 1.4$ Gyr and there is no substantial population in the tails of the line ratio distribution, if the turnoff process involves one of these two emission lines substantially disappearing, we have seen no examples and therefore the turnoff process must require no more than tens of Myr to complete.  Alternatively, this places no limits on the timescale for a low-redshift turnoff process involving no changes in these line ratios.

At $1.8 < z < 2.0$, the Mg{\small II} to C{\small IV} line ratio changes with mass, with the Mg{\small II} line being stronger at high black hole mass.  One possibility is that massive black holes tend to have strongly asymmetric C{\small IV} lines\cite{Shen2008}, and therefore the line flux is underestimated by a Gaussian fit.  However, the C{\small IV} line has a higher ionization potential.  As a result, perhaps we could interpret Figure \ref{fig:10} as showing that massive black holes have a different ionizing continuum than less massive black holes near $z \sim 2.0$.  This might give us a hint as to the turnoff mechanism for massive black holes at $z \sim 2.0$.

It is curious that the Mg{\small II} to C{\small IV} ratio changes with mass $1.8 < z < 2.0$ while at $0.6 < z < 0.8$ the H$\beta$/Mg{\small II} ratio is not affected.  Because the three lines come from different ionization potentials, this apparent discrepancy might indicate that the turnoff process acts differently on the broad-line region and the ionizing continuum.  Alternatively, it might indicate that the turnoff mechanism is changing between these two redshift ranges.  Additional spectroscopy in the infrared and ultraviolet is possible and will be able to determine which interpretation is correct.

\section{A Synchronization Puzzle}
\label{sec:synch}

In \S~\ref{sec:permanent}, it was shown that the number density of quasars of a given $M_{BH}$ declines with a mass-dependent e-folding time between 0.7 and 3 Gyr for quasars with an $M_{BH}$ of  $10^9 M_\odot - 10^{10} M_\odot$ (Table \ref{table:decays}).  The SDSS catalog includes much of the Northern hemisphere and so includes quasars almost diametrically opposed in the sky.  At a redshift of 2, they lie in host galaxies that had not been causally connected since inflation, while even at much lower redshift it is difficult to believe galaxies a few Gpc apart would strongly influence each others' development.  Yet, quasars with $M_{BH} \sim 10^{10} M_\odot$ in such galaxies turn off synchronously to within 700 Myr.

Could this be resolved if virial mass estimates are poorly calibrated at high mass?  While the details of quasar turnoff do rely upon high-mass objects, there is a similar synchronization in the sub-Eddington boundary (Paper I)\nocite{Steinhardt2009}.  This boundary restricts quasars to a luminosity below $L_{Edd}$ at most masses and every redshift, with the restriction stronger with increasing mass.  However, for every mass there is a redshift above which there is a substantial population accreting near $L_{Edd}$.  This again presents a synchronization problem; for example, a substantial population of quasars with $M_{BH} \sim 10^8$ are at Eddington at $z \sim 0.6$, but none are at Eddington at $z \sim 0.4$, less than 2 Gyr later.  So it is not just turnoff that is synchronized for quasars at a given $M_{BH}$ but rather most of their luminous accretion.  Because the location of the sub-Eddington boundary declines in luminosity over time at a given mass, its sharpness is further evidence of this universal synchronization of quasar accretion.

Each quasar lies within a host galaxy, and it is widely assumed that the dynamics of quasar accretion are controlled by the dynamics of its host, since the host galaxy appears to be the only fuel source available.  Turnoff, then, would occur when the galaxy no longer can sufficiently fuel its central black hole, a condition presumably related to the evolution of the host galaxy.  Further evidence for co-evolution of the quasar and the host galaxy comes from the black hole mass - stellar velocity dispersion ($M_{BH} - \sigma$) relation \cite{msigma1,msigma2} as well as because quasars of a given mass tend to also lie in host galaxies of a common mass \cite{msigma1,msigma2}. 

However, host galaxies of a given mass are not believed to be as well synchronized as required by quasar accretion or turnoff.  The Millenium simulation \cite{Springel2005} continues to produce new galaxies at low redshift, while galaxy merger rates are believed to only decline by a factor of $\sim 10$ between a redshift of 2 and today \cite{Hopkins2009b}, equivalent to a characteristic e-folding timescale of $\sim 5$ Gyr.  In short, if the dynamics of quasars are only as synchronized as those of their host galaxies, our current theoretical understanding of galactic dynamics is incompatible with the data presented here.

\section{Discussion}
\label{sec:discussion}

Here and in Paper I \nocite{Steinhardt2009}, we have considered what the quasar locus in the mass-luminosity plane can teach us about quasar accretion and turnoff.  Quasars have been modeled as very simple objects, in part because there has not yet been sufficient data to require more complex assumptions.   In Paper I, we falsified the assumption that quasars at all combination of mass and redshift were near their Eddington luminosity while accreting luminously.  With a combination of the large quasar catalog from the Sloan Digital Sky Survey (SDSS) and virial mass estimation, we can now evaluate several other common assumptions.

It is widely assumed that the dynamics of quasar turnoff are closely linked to the dynamics of the host galaxy.  However as shown in \S~\ref{sec:synch}, accretion rate and turnoff are highly synchronized amongst quasars of a common mass.  Either galaxies must be much more highly synchronized than currently believed or this synchronization must have an alternate source.  Galaxies continue to merge and virialize at later times than their quasars are predominantly turning off, and the decline in galaxy merger rates is believed to be much more gradual than that observed for quasars.  So, why are quasars so well synchronized?  One possibility is that the seeding mechanism comes not from the dynamics of their host galaxy but rather from some mechanism with a cosmic clock.  The $\sim 700$ Myr timescale for quasar turnoff at $M_{BH} \sim 10^{10} M_\odot$ corresponds to the Hubble time at $z \sim 8$, and extrapolating the slope of the e-folding time as a function of mass (\S~\ref{sec:permanent}) to the highest-mass quasars would yield the Hubble time at $z \sim 20$.  It is plausible that the seeds of these massive quasars turn on during the epoch of reionization and, if the seeding mechanism is one that yields a nearly-fixed quasar life span, turnoff would be synchronized to roughly the observed precision.  This would require a seeding mechanism only weakly linked to the dynamics surrounding galaxy, but rather triggered by cosmic dynamics.

It is also widely assumed that most quasars are near Eddington while accreting luminously.  The presence of a sub-Eddington boundary at which high-mass quasars cannot reach their full Eddington luminosity directly contradicts this claim and slows the maximum allowed growth rate.  However the boundaries of the quasar locus are very sharp, and quasars of a given mass are highly synchronized in both the maximum luminosity they can attain and in their turnoff time.  In the absence of techniques for mass estimation of high-redshift quasars, it was widely assumed that a segregation of quasars by mass was equivalent to a segregation by luminosity.  Since individual quasars are variable, this cannot be entirely true, but there is certainly a correlation between quasar mass and luminosity as evidenced by the quasar distribution in the mass-luminosity plane (Figure \ref{fig:allzcontour}).  Further, the high-mass, low-luminosity boundary suggests that even if high-mass quasars cannot reach $L_{Edd}$, luminous accretion is rare below a few percent of Eddington as well.  This also implies that the large Milliquas sample \cite{Milliquas} of $\sim 10^6$ new SDSS quasars selected photometrically but not bright enough for SDSS spectroscopy consists predominantly of quasars at lower $M_{BH}$ than those in the SDSS spectroscopic quasar catalog rather than a mixture of quasars at low $M_{BH}$ and quasars at higher $M_{BH}$ and very low Eddington ratio.

Finally, it is widely believed that a typical SMBH will have many periods of luminous accretion separated by dormant periods.  In part, this is because the Salpeter time for black hole accretion at $L_{Edd}$ is only $\sim 10^8$ yr \cite{Natarajan1998}, and a quasar could not grow at $L_{Edd}$ for several Gyr.  However, the sub-Eddington boundary restricts more massive quasars at every redshift to $0.2~L_{Edd}$ or below (with the peak number density below $0.1~L_{Edd}$), raising the possibility that rather than accreting luminously for less than 1 Gyr, quasars might accrete luminously, but sub-Eddington, in one long, continuous burst for several Gyr from when they first turn on until turnoff.  Certainly the sub-Eddington boundary requires that quasars accrete luminously far longer than previously believed.  

When quasars of a given mass no longer reach a given Eddington ratio, the sub-Eddington boundary never moves back to a higher luminosity.  When quasars of a given mass begin to decline in number density, their turnoff is synchronized and sharp, and the population does not reappear at lower redshift.  While this is insufficient evidence to conclude that turnoff is a permanent process for any individual quasar, a quasar lifecycle consisting of one long period of accretion followed by permanent turnoff is consistent with all of these observations.

As shown in \S~\ref{sec:signatures}, the sharpness of the high-mass, low-luminosity boundary (HMLLB) appears to require either that virial mass estimation is wrong near the HMLLB or that quasars at the HMLLB undergo a rapid transition to an object with a very different spectral energy distribution.  One possibility would be transition to RIAF accretion \cite{Narayan2008}.  However, the HMLLB does not occur at a constant Eddington ratio as accretion disk theory would suggest, but rather has a sub-linear dependence on mass.  The HMLLB is also evolving with redshift:  the HMLLB for high-mass quasars is at a higher redshift than lower-mass quasars.

In this paper, we have shown several new puzzles presented by the quasar mass-luminosity plane in addition to the sub-Eddington boundary discussed in Paper I.  These new puzzles which cannot be explained by selection effects are further evidence that supermassive black hole accretion and turnoff is more complex than previously believed.  Rather, many of the common assumptions made in quasar toy models are poor matches to the quasar locus in the $M-L$ plane.  These new boundaries indicate both that a more nuanced quasar model is required and that the quasar mass-luminosity plane now contains enough detail that it might be able to provide strong tests for such new models.

The authors would like to thank Mihail Amarie, Doug Finkbeiner, Lars Hernquist, John Huchra, and Michael Strauss for valuable comments.  This work was supported in part by Chandra grant number 607-8136A.

\end{document}